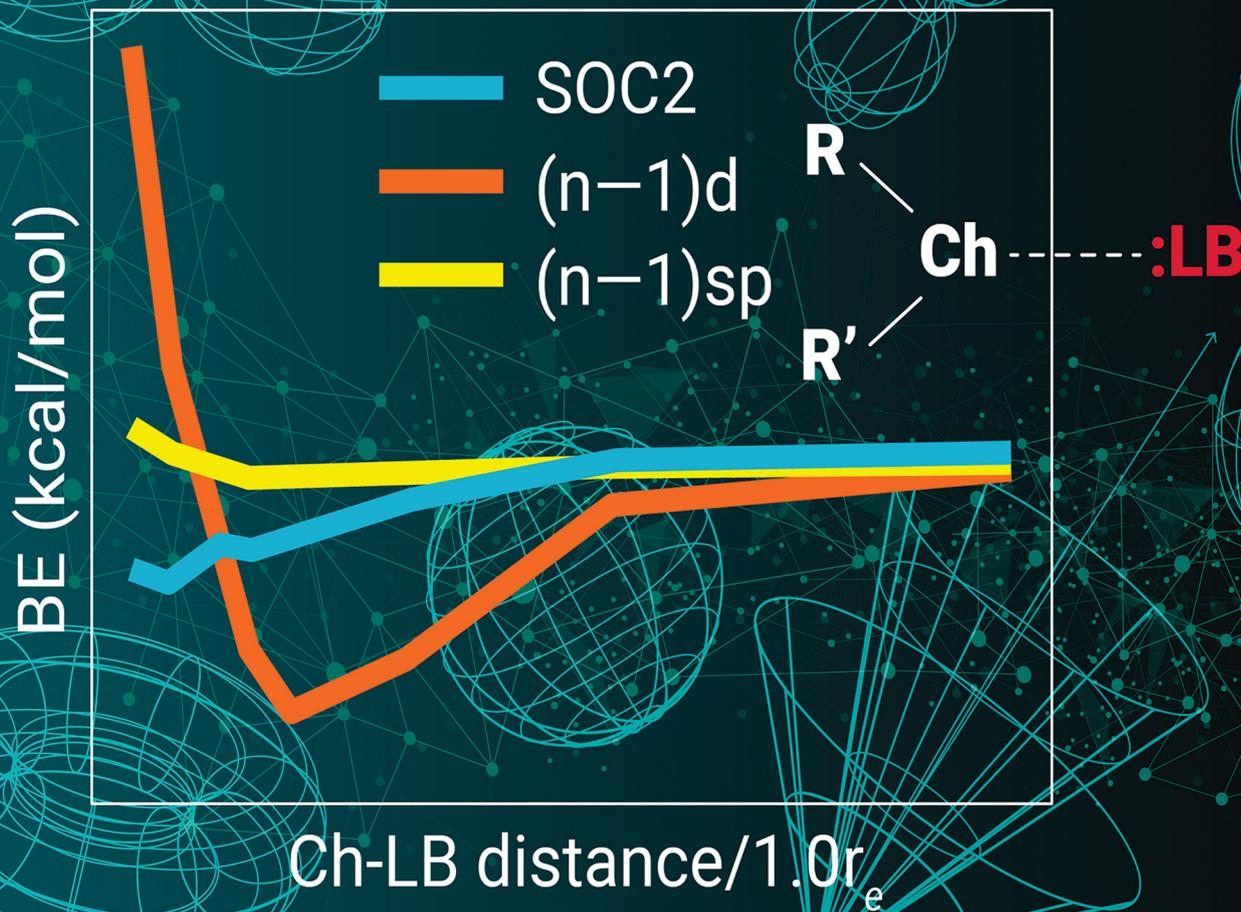

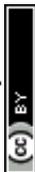

# Exploring the influence of ($n - 1$)d subvalence correlation and of spin–orbit coupling on chalcogen bonding†

Nisha Mehta *‡ and Jan M. L. Martin

This article presents a comprehensive computational investigation into chalcogen bonding interactions, focusing specifically on elucidating the role of subvalence ($n - 1$)d and ($n - 1$)sp correlation. The incorporation of inner-shell ($n - 1$)d correlation leads to a decrease in interaction energies for chalcogen-bonded systems (at least those studied herein), contradicting the observations regarding halogen bonding documented by Kesharwani *et al.* in *J. Phys. Chem. A*, **2018**, 122 (8), 2184–2197. The significance of ($n - 1$)sp subvalence correlation appears to be lower by an order of magnitude. Notably, among the various components of interaction energies computed at the PNO-LCCSD(T) or DF-CCSD levels, we identify the PNO-LMP2 or DF-MP2 component of the ($n - 1$)d correlation as predominant. Furthermore, we delve into the impact of second-order spin–orbit coupling (SOC2) on these interactions. The SOC2 effects appear to be less significant than the ($n - 1$)d correlation; however, they remain non-trivial, particularly for Te complexes. For the Se complexes, SOC2 is much less important. Generally, SOC2 stabilizes monomers more than dimers, resulting in reduced binding of the latter. Notably, at equilibrium and stretched geometries, SOC2 and ($n - 1$)d destabilize the complex; however, at compressed geometries, they exhibit opposing effects, with ($n - 1$)d becoming stabilizing.

## 1 Introduction

Non-covalent interactions exert a crucial influence on the physical and chemical properties of diverse systems.[1,2] They are pivotal in processes such as protein folding, ligand interactions, packing and stacking arrangements, organocatalysis, supramolecular chemistry, and conformational stability, underscoring their significance across a spectrum of scientific disciplines.[3,4] A molecule featuring a chalcogen atom, such as sulfur, selenium, or tellurium, can partake in a multitude of non-covalent interactions, exemplified by phenomena such as chalcogen bonding, electrostatic interactions, and hydrogen bonding. Chalcogen bonding is a type of non-covalent interaction that involves the interactions between a chalcogen atom (chiefly S, Se, or Te) and a Lewis base or electron-rich region in the neighboring molecule. In chalcogen bonding, a chalcogen atom functions as an electrophile, while the interacting partner contributes electron density to facilitate the interaction. These interactions exhibit a pronounced directional character. The importance of chalcogen bonding interactions lies in their role as a unique and versatile class of noncovalent interactions, influencing molecular recognition, supramolecular assembly, and crystal engineering, with potential applications in drug design, materials science, and catalysis.[5–17]

Numerous systems engage in chalcogen bonding *via* the σ-hole; nonetheless, an alternative avenue involves the π-hole. This distinctive π-hole constitutes a positively charged domain, positioned orthogonally to a planar π-framework. This positive region demonstrates an affinity for interacting with electron donors. Illustrative instances encompass interactions like $SO_3 \cdots H_2O$,[18–20] $SO_3 \cdots NH_3$,[21–23] $(SO_3)_n \cdots H_2CO$,[24,25] $(SO_3)_n \cdots CO$,[26] and $(SO_3)_n \cdots (CO)_m$,[26] where $n$ = 1,2.

Computational methodologies, such as *ab initio* calculations, play a pivotal role in the examination of chalcogen bonding interactions. It enables precise predictions of molecular structures, energetics, and electronic properties, yielding invaluable insights that might pose challenges or prove unattainable through experimental means. Furthermore, the results can function as a reference point for refining less expensive computational approaches, like DFT and force field methods. Nonetheless, to conduct a thorough benchmark study, it is imperative to accumulate statistics from a sufficiently extensive array of calculations. Illustrative examples of such comprehensive datasets include the G$n$ test sets,[27–30] database 2015B,[31] MGCDB84 (Main Group Chemistry Data Base, 84 subsets) of the Berkeley group,[32] and the Grimme and Goerigk groups' GMTKN55[33] dataset (general main-group thermochemistry, kinetics, and noncovalent interactions, 55 problem sets), as well as the latter's predecessors GMTKN24[34] and GMTKN30.[35] These extensive databases encompass a diverse array of benchmark sets, including directional non-covalent interactions such as hydrogen and halogen bonding. While prior research extensively explored halogen bonding interactions, as detailed in the references, it is noteworthy that the aforementioned datasets do not encompass complexes representing chalcogen bonding interactions. To address this gap, one of us and coworkers previously developed the CHAL336 benchmark[36] comprising 336 chalcogen-bonded dimers, hitherto the largest and most accurate of its kind.

*Department of Molecular Chemistry and Materials Science, Weizmann Institute of Science, 7610001 Rehovot, Israel. E-mail: nisha.mehta@weizmann.ac.il*

† Electronic supplementary information (ESI) available: Spreadsheet containing RMSD at various levels of theory, results of SAPT and other NCI indices; Cartesian coordinates (in.xyz format) of all structures; interaction energies at various levels of theory (in.txt format). See DOI: https://doi.org/10.1039/d4cp01877j

‡ Present address: School of Chemistry, The University of Melbourne, VIC 3010, Australia.

As we delve into the intricacies of chalcogen bonding, our focus narrows down to a crucial and often overlooked aspect—the importance of subvalence $(n-1)$d and $(n-1)$sp correlation contributions to chalcogen bonding interactions. For other types of noncovalent interactions, especially involving lighter elements, inner-shell correlation is routinely neglected—this is easy to justify for elements like O and even S, where the core-valence gaps are 19.4 and 5.8 Hartree, respectively. For Se and Te, however, the outer-core $(n-1)$d orbitals lie just 1.8 and 1.3 Hartree, respectively, below the valence shell—even smaller than in the adjacent halogens Br and I, for which it has previously been shown[37] that $(n-1)$d correlation contributes quite significantly to the interaction energy, particularly at shorter distances.

However, if we include core-valence effects, it behooves us to also consider whether other contributions could not be of similar importance. What comes to mind in particular is spin–orbit coupling (SOC), a quantum phenomenon resulting from the interaction between an electron's spin and the magnetic field generated by its orbital motion around the nucleus. SOC can exert a significant influence on the electronic and magnetic properties of materials, particularly those containing heavy elements with large atomic numbers. Closed-shell systems are not subject to first-order SOC, but they may be stabilized by second-order SOC (SOC2: see, e.g.,[38–41]). For instance, De Jong and coworkers[42] found for HBr, Br$_2$, HI, and I$_2$ 2nd order SOC stabilizations of {0.1, 0.4, 0.5, 2.0} kcal mol$^{-1}$, respectively. An experimental manifestation of SOC2 is the zero-field splitting (ZFS) in the X$^3\Sigma^-$ ground states of chalcogen diatomics: see, e.g., Table 6 in ref. 43, where one finds ZFS of 23.5 cm$^{-1}$ for S$_2$, 510 cm$^{-1}$ for Se$_2$, and 1975 cm$^{-1}$ for Te$_2$. (That is, 0.067, 1.458, and 5.647 kcal mol$^{-1}$, respectively—note a rough $Z^4$ dependence on the atomic number $Z$ as conjectured in ref. 44 and noted, in a solid state physics context, in ref. 45). As such numbers are actually in the same energetic range as the subvalence (outer-core) correlation effects we are considering here, we ought to at least attempt to gauge the importance of SOC2 for chalcogen bonding interactions, and compare its importance with that of outer-core correlation.

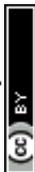

## 2 Computational details

We have selected TeO$_3$···TeHF, SeO$_3$···SeHF, and SO$_3$···SHF as exemplars of chalcogen bonding interactions involving Te, Se, and S, respectively. In this study, the reference geometries for TeO$_3$···TeHF, SeO$_3$···SeHF, and SO$_3$···SHF were acquired from ref. 36 and employed without additional optimization. The original 1.0$r_e$ structure had undergone optimization using the PW6B95[46]-D3(BJ)[47] method with a def2-TZVPD[48–50] basis set. In our current study, we determined the remaining structures through stretching and compressing the intermonomer distance (with frozen monomer geometries) by scaling factors of {0.80, 0.85, 0.90, 0.95, 1.0, 1.05, 1.10, 1.25, 1.50, 2.0} using an in-house Python script. The XYZ coordinates for all 30 structures can be found in the ESI.† Initially, we included compressed distances of 0.80$r_e$ and 0.85$r_e$; however, we subsequently opted to omit them due to the strong intermonomer repulsion at such compressed geometries. Consequently, the statistics presented in this manuscript pertain to distances of {0.90, 0.95, 1.0, 1.05, 1.10, 1.25, 1.50, 2.0}, totaling 24 dimers. In response to one of the reviewer's requests, we have also included TeHF···NH$_3$, SeHF···NH$_3$, and SHF···NH$_3$ in our study. The geometries of these molecules were optimized using the PW6B95[46]/def2-TZVPP[48,49] level of theory with ORCA 5.0.2[51–53] for the optimization process. Additionally, we obtained stretched and compressed intermonomer distances (while keeping the monomer geometries fixed) by applying scaling factors of {0.90, 0.95, 1.0, 1.05, 1.10, 1.25, 1.50, 2.0} through the aforementioned Python script. The corresponding XYZ coordinates are provided in the ESI.†

The computations were conducted on the "ChemFarm" HPC cluster of the Faculty of Chemistry at the Weizmann Institute of Science. All *ab initio* calculations, both canonical and localized, were performed using MOLPRO 2023.2.[54] Specifically, we employed a combination of Dunning correlation-consistent cc-pV$n$Z (where $n$ = D, T, Q, 5) basis sets[55] for hydrogen atoms, along with their corresponding augmented counterparts, aug-cc-pV$n$Z,[56] for non-hydrogen atoms other than the chalcogens sulfur, for which we employed aug-cc-pWCV$n$Z,[57,58] and selenium and tellurium, for which we utilized aug-cc-pWCV$n$Z-PP.[48,59,60] (In this context, we would like to mention ref. 61 and references therein, which recommended the incorporation of aug-cc-pWCV$n$Z basis sets for the accurate treatment of core-electron correlation.)

Our canonical calculations were limited to the density fitting CCSD (coupled cluster with iterative singles and doubles[62]) level as implemented in MOLPRO. For the higher CCSD(T) level,[63] we employed localized orbital CCSD(T) approximations, and specifically the PNO-LCCSD(T) method (pair natural orbital localized coupled cluster, see ref. 64 for a review), and using the DomOpt = Normal, Tight, and vTight threshold combination as detailed in the MOLPRO 2024 online manual[65]—which for Tight differ slightly from the original values given in Ma and Werner.[66]

Basis set extrapolation was carried out by means of the two-point extrapolation formula $E_L = E_\infty + B/L^\alpha$, where $L$ denotes the maximum angular momentum present within the basis set and $\alpha$ represents the exponent associated with the level of theory and basis set pair. Equivalently,[67] this can be written in the Schwenke[68] form $E_\infty = E_L + A_L(E_L - E_{L-1})$, if the Schwenke extrapolation coefficient $A_L = \left(\left(\frac{L}{L-1}\right)^\alpha - 1\right)^{-1}$. (For a discussion of the equivalence relations between the various two-point extrapolation formulas, see ref. 67).

For the HF, MP2, CCSD, and (T), our {T,Q}Z Schwenke coefficients are 0.415, 0.915, 0.700, and 0.730, respectively, taken from ref. 67 and 69. Likewise, for the {Q,5}Z extrapolation, the corresponding values are 0.528, 1.208, 0.930, and 0.810. It should be noted that we employed identical extrapolation exponents for both the localized and canonical methods, as they should converge to the same result.

Spin–orbit coupling calculations were performed at the STEOM-CCSD[70] level as implemented in the ORCA software package (version 5.0.2).[51–53] The zeroth order regular approximation (ZORA) method was used to account for scalar relativistic effects.[71] The aug-cc-pVTZ-DK basis set[58,72–74] was employed, in conjunction with the auxiliary basis sets SARC/J,[75–79] and def2-TZVPP/C[80,81]. Spin–orbit integrals were evaluated within the RI-SOMF(1X) approximation.[82] In addition, we also computed spin–orbit coupling calculations at the CAM-B3LYP[83]-/aug-cc-pVQZ-DK level using the time-dependent density functional theory (TD-DFT) module implemented in the ORCA software package.

In addition to the Boys–Bernardi counterpoise corrections[84] and the original "raw" (uncorrected) values, we also employ the mean of both (referred to as "half-CP"), a practice rationalized by Sherrill *et al.*[85] and by Brauer *et al.*[86]

Finally, symmetry-adapted perturbation theory (SAPT, see ref. 87 for a review) calculations were performed employing the SAPT module[88] within the PSI4 software framework.[89]

## 3 Results and discussion

The complete list of the 24 chalcogen-bonded systems, together with our best values for the interaction energies obtained in the present work, is outlined in the final two columns of Table 3. It is crucial to note that throughout this manuscript, positive interaction energy or BE values signify dimer stabilization, as we have reported the negative of the difference between the complex and monomer energies. Due to hardware limitations, we were unable to compute interaction energies utilizing the haWCV5Z basis set at the canonical CCSD level, nor could we incorporate (T). Therefore, our canonical calculations were limited to the DF-CCSD/haWCV{T,Q}Z level. However, we were able to conduct computations using the 5Z-sized basis set through the localized PNO-LCCSD(T) scheme. We have selected the PNO-LCCSD(T) approach, opting for the (Tight, {Q,5}Z) half-CP as our optimal reference level. Localized methods such as PNO-LCCSD(T) (tight, {Q,5}Z) serves as a suitable alternative when canonical references are not feasible, as recommended in ref. 90–93. Both methods converge to the same values at the complete basis set limit. Although we will delve into this more thoroughly in the following section, it is crucial to note that our reference level (*i.e.*, PNO-LCCSD(T) (Tight, {Q,5}Z) half-CP exhibits an RMSD of just 0.046 kcal mol$^{-1}$ when compared to both the raw and full-CP counterparts (Table 1). Needless to say, at the true CBS limit, the difference between raw and CP-corrected values should be zero: a significant discrepancy between raw and CP in a CBS extrapolation suggests that the underlying calculations are inadequate or that the extrapolation procedure is problematic or both.

Table 1 The effect of incorporating counterpoise correction on the different components of total interaction energies computed at the PNO-LCCSD(T)(tight, haWCV{Q,5}Z) level. The RMSD values presented are for the raw/full-CP counterparts in comparison to the half-CP. (A) shows the statistics for 'valence + subvalence $(n - 1)$d' electron correlation, while (B) shows the corresponding values for the 'subvalence $(n - 1)$d component only. The full statistics across all basis sets (TZ, QZ, {T,Q}Z, 5Z, {Q,5}Z) as well as different DomOPT settings (Default, Tight, vTight) and canonical CCSD results for TZ, QZ, {T,Q}Z basis sets are reported in the ESI

|  | (A) | | | | (B) | |
|---|---|---|---|---|---|---|
|  | Te | Se | S | All | Te | Se |
| HF | 0.012 | 0.010 | 0.006 | 0.010 | | |
| PNO-LMP2 | 0.027 | 0.018 | 0.026 | 0.024 | 0.027 | 0.005 |
| PNO-LCCSD | 0.055 | 0.038 | 0.006 | 0.039 | 0.017 | 0.002 |
| (T) | 0.011 | 0.007 | 0.002 | 0.007 | 0.003 | 0.001 |
| PNO-LCCSD(T) | 0.065 | 0.044 | 0.007 | 0.046 | 0.020 | 0.003 |
| CCSD-MP2 | 0.029 | 0.023 | 0.025 | 0.026 | 0.010 | 0.003 |
| CCSD(T)-MP2 | 0.040 | 0.029 | 0.026 | 0.032 | 0.007 | 0.002 |

We performed three distinct sets of computations: (i) solely considering valence electrons, (ii) including valence and $(n-1)$d electrons, and (iii) encompassing valence, $(n-1)$d, and $(n-1)$sp electrons. Unless otherwise specified, the statistical analyses presented in subsequent section pertains to 'valence + $(n-1)$d electrons'.

### 3.1 Some observations on the counterpoise corrections for the chalcogen bonding interactions

First we wish to establish how basis set superposition error (BSSE) behaves for different components of the interaction energy (SCF, MP2 correlation, and post-MP2 correction) and obtain interaction energies with basis sets large enough that BSSE has been reduced to insignificance. It has been argued numerous time in the literature (perhaps the first time by Helgaker and coworkers[94]) that BE$_{raw}$ and BE$_{CP}$ tend to converge to the CBS limit from opposite directions: hence, BE$_{half-CP}$ = (BE$_{raw}$ + BE$_{full-CP}$)/2 naturally suggests itself as the optimal choice (see ref. 85, 86 and references therein). When observing smaller RMSDs for CP-uncorrected results, this is mainly due to error compensation between BSSE (which causes overbinding) and intrinsic basis set insufficiency (IBSI, which leads to underbinding).[86] In such cases, this suggests the need for larger basis sets. It has been shown in the past (*e.g.*, ref. 69 and 85) that even a 5Z-sized basis set may still be too far from CBS for an MP2 or coupled cluster calculation for its BSSE to outweigh its IBSI; hence, applying half-CP correction on top of {Q,5}Z extrapolation appears to be more appropriate.

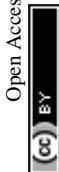

**SCF and MP2 components.** First, let us delve into the HF level. In our computations using HF/haWCV{Q,5}Z pantry, the impact of counterpoise corrections appears minor, albeit non-negligible: the RMSD between half-CP versus full-CP and raw data stands at 0.010 kcal mol$^{-1}$ (Table 1). Analysis of Table S1 (ESI†) indicates that at the haWCV5Z and haWCVQZ levels, all three variants—raw, half-CP, and full-CP—yield RMSD values of approximately 0.03 and 0.08 kcal mol$^{-1}$, respectively. Furthermore, the {T,Q}Z half-CP demonstrates a considerable 0.225 kcal mol$^{-1}$ RMSD disparity from the reference value, which is quite substantial. At the TZ level, the RMSD becomes more pronounced, reaching 0.393 kcal mol$^{-1}$ for the half-CP.

Thus, we observed a slower-than-usual convergence of basis sets for the SCF component. Although no systematic trend is evident, Table S1 (ESI†) suggest that basis set convergence is slower for sulfur-based complexes, followed by tellurium and selenium.

Let us next consider the MP2 correlation component. At the PNO-LMP2 (Tight, {Q,5}Z) level, the RMSD of half-CP with both CP-corrected and uncorrected interaction energies is just 0.024 kcal mol$^{-1}$. {T,Q}Z extrapolation seems to yield lower RMSDs than 5Z. As an example, 5Z yields RMSD of 0.496 kcal mol$^{-1}$, in comparison to just 0.073 kcal mol$^{-1}$ for {T,Q}Z, this is even the case for post-MP2 corrections as well, which we will be discussing next. It appears that {T,Q}Z extrapolation may potentially yields superior results compared to utilizing the 6Z basis set. For example, in a very recent study on basis set extrapolations,[95] it has been shown that for the W4-17 thermochemical benchmark,[96] the RMSD computed for BSSEs at the CCSD/haV$n$Z + d level are {3.94, 1.74, 0.77, 0.31} kcal mol$^{-1}$ for $n$ = {T,Q,5,6}, in contrast to 0.27 kcal mol$^{-1}$ for haV{T,Q}Z and 0.13 kcal mol$^{-1}$ for haV{Q,5}Z.

We observed an increase in RMSDs upon incorporating counterpoise correction. Sorting by RMSD may suggest that opting for the "raw" data is optimal for the majority of basis sets. For instance, even for a 5Z-sized basis set, the RMSD increases from 0.326 kcal mol$^{-1}$ to 0.496 and 0.667 kcal mol$^{-1}$ upon incorporating half and full-CP corrections, respectively. However, this does not imply that one should avoid using counterpoise correction. Rather, as explained at length by Sherrill and coworkers[85] for orbital-based calculations, and by Brauer et al.[86] for explicitly correlated ones, such 'right trends for the wrong reason' reflect that 'counterpoise-uncorrected' interaction energies suffer from error compensation between BSSE and IBSI. In such cases, larger basis sets are indicated.

Furthermore, canonical DF-MP2 calculations using the haWCV{T,Q}Z extrapolation with half-counterpoise correction yield an RMSD of only 0.059 kcal mol$^{-1}$ from the localized reference level (Table S1 in ESI†). This suggests that both our localized and canonical outcomes are converging towards the same basis set limit, affirming that any observations made with the localized scheme are not merely artifacts of its usage.

**Post MP2 correction.** Next, let us delve into the post-MP2 correlation components. Beginning with the CCSD-MP2 differences, at the reference level (i.e., PNO-LCCSD – PNO-LMP2/(tight, haWCV{Q,5}Z) half-CP), the difference between half-CP values and CP-corrected and uncorrected interaction energies is merely 0.026 kcal mol$^{-1}$ RMSD.

It is noteworthy that the CCSD-MP2 term exhibits a more rapid basis set convergence compared to the MP2 term discussed earlier (refer to Table 1 and Table S1, ESI†). Interestingly, this finding contradicts the trend observed for atomization energies, as documented by Ranasinghe and Petersson.[97]

At the {T,Q}Z/Tight level, the RMSD for half-CP is just 0.046 kcal mol$^{-1}$, slightly larger than for full-CP (0.036 kcal mol$^{-1}$) but lower than for uncorrected results (0.080 kcal mol$^{-1}$). RMSD values at the canonical DF-CCSD – DF-MP2/haWCV{T,Q}Z are equally small (0.124, 0.079 and 0.042 kcal mol$^{-1}$, respectively, for the full-CP, half-CP and 'raw' results).

What about the CCSD(T)-MP2 component? At the reference level [PNO-LCCSD(T) – PNO-LMP2/(tight, haWCV{Q,5}Z) half-CP], the discrepancy between half-CP values and CP-corrected and uncorrected interaction energies is merely 0.032 kcal mol$^{-1}$ RMSD—comparable in magnitude to what we observed for the CCSD-MP2 (i.e., PNO-LCCSD – PNO-LMP2/(tight, haWCV{Q,5}Z) half-CP)) component. While it is customary in noncovalent interactions to generally consider CCSD(T)-MP2 together without further separation, in this study, it may be worthwhile to delve into the (T) component separately. This is particularly relevant for the upcoming section, where we will explore that the incorporation of subvalence ($n$ − 1)d electrons—central to the focus of this manuscrip—may not significantly impact the (T) component. It's worth noting that the basis set convergence of the (T) component is faster compared to MP2 and CCSD-MP2 components. Uncorrected and counterpoise-corrected (T)/haWCV{Q,5}Z results are practically identical, with an RMS difference of only 0.007 kcal mol$^{-1}$. (T)/{T,Q}Z extrapolation with tight settings and half-CP yields RMSD of 0.015 kcal mol$^{-1}$. Once more, for most basis sets without CBS extrapolation, counterpoise corrections seem to be disadvantageous.

In summary, for chalcogen bonding interactions, CP correction generally worsens the statistical outcomes even with basis sets as large as 5Z—owing to error cancellation between basis set superposition error and intrinsic basis set insufficiency, highlighting the necessity of employing large basis sets, particularly for the MP2 component, when studying such interactions. It seems that employing CBS extrapolation and applying CP correction atop it represents the best strategy. Furthermore, {T,Q}Z extrapolation surpasses the performance of the 5Z basis set.

### 3.2 Effect of ($n$ − 1)d and ($n$ − 1)sp subvalence electron correlation

To examine core-valence correlation contributions, we conducted three sets of calculations for a given level of theory: (i) valence electrons only, (ii) valence + ($n$ − 1)d electrons, and (iii) valence + ($n$ − 1)d + ($n$ − 1)sp electrons. The interaction energy contributions calculated are provided in Table 2 and ESI.†

At the reference level [i.e., PNO-LCCSD(T)/(tight, hAWCV-{Q,5}Z) half-CP], the inclusion of ($n$ − 1)d subvalence correlation destabilizes the TeO$_3$···TeHF complex by 0.386 kcal mol$^{-1}$.

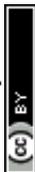

As the $TeO_3 \cdots TeHF$ distance is increased, this destabilization effect intensifies, reaching a maximum at $1.10r_e$, where $(n − 1)d$ subvalence correlation contributes 1.119 kcal mol$^{-1}$ to the destabilization. Beyond this point, the impact of $(n − 1)d$ subvalence correlation diminishes, with a contribution of only 0.021 kcal mol$^{-1}$ at $2.0r_e$, still contributing to the overall decrease in interaction energy. For the $SeO_3 \cdots SeHF$ complex, the maximum destabilization effect is observed at $1.05r_e$, with a value of 0.636 kcal mol$^{-1}$. The RMSD values between the 'valence only' and 'valence + $(n − 1)d$' calculations are 0.891 kcal mol$^{-1}$ for Te complexes and 0.428 kcal mol$^{-1}$ for Se complexes.

Furthermore, the incorporation of $(n − 1)sp$ correlation effects in Te complexes slightly increases the destabilization. The RMSD values between 'valence + $(n − 1)d$' and 'valence + $(n − 1)spd$' are 0.067 kcal mol$^{-1}$ and 0.191 kcal mol$^{-1}$ for Te and Se complexes, respectively, indicating that the impact of $(n − 1)sp$ subvalence correlation is an order of magnitude lower compared to the effects observed for $(n − 1)d$ subvalence correlation.

Before we delve into more detail on the effect of $(n − 1)d$ inner shell correlation, we first should explore its basis set convergence, as shown in Table S4 in the ESI.† A similar table for $(n − 1)sp$ is also provided in the ESI.† Te complexes exhibit significantly slower basis set convergence for the $(n − 1)d$ inner shell correlation component compared to Se complexes, as discussed above. While exploring how BSSE behaves for the different components of the $(n − 1)d$ contribution to the total interaction energy (MP2 and post MP2-correction), we again noticed that for all tripleZ–5Z basis sets, CP correction mostly degrades the statistics. However, for CBS extrapolated values (where the lion's share of basis set errors is already taken care of), CP correction does help. Once again, CP-uncorrected data benefits from error cancellation between BSSE and IBSI as their effects are in opposite directions. Furthermore, the (T) component of the $(n − 1)d$ correlation displays the most rapid basis set convergence.

Table 2 presents the $(n − 1)d$ component of the total interaction energies calculated at the localized PNO-LCCSD(T) and canonical DF-CCSD/haWCV{T,Q}Z levels for all 24 systems studied herein. (The corresponding table for $(n − 1)sp$ can be found in the ESI†). Among different components of PNO-LCCSD(T) (or DF-CCSD), PNO-LMP2 (or DF-MP2) contributes the lion's share. The RMSD between valence and valence + subvalence $(n − 1)d$ for PNO-LMP2 is 1.135 kcal mol$^{-1}$. This discrepancy is not an artifact of the localized scheme, as we also obtained an RMSD of 1.154 kcal mol$^{-1}$ at the MP2/haWCV{T,Q}Z level. The CCSD-MP2 component displays around one-fourth of the RMSD observed for MP2. Furthermore, the (T) component of the $(n − 1)d$ correlation is extremely small, measuring only 0.039 kcal mol$^{-1}$. These findings are crucial, as they might prove useful in future work. Keeping the calculations including $(n-1)$d correlation down to the MP2 level leads to substantial savings in CPU time and memory/mass storage overhead.

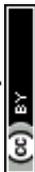

Table 2 The $(n − 1)d$ component of total interaction energies (in kcal mol$^{-1}$), computed at the PNO-LCCSD(T)(tight, haWCV{Q,5}Z) half-CP and CCSD/hAWCV{T,Q}Z half-CP levels. It includes breakdowns of the $(n − 1)d$ contributions into MP2, CCSD, (T), CCSD-MP2 and CCSD(T)-MP2 levels. Heatmapping is employed to illustrate the contribution's impact on dimer stability, ranging from blue (stabilizing) to red (destabilizing). Additionally, the lower panel of the table presents RMSD and MAD values (in kcal mol$^{-1}$). Here, the heat mapping indicates deviations, ranging from red (for larger deviations) transitioning to yellow and green (for smaller deviations)

| | PNO-LCCSD(T)/(tight, haWCV{Q,5}Z) half-CP | | | | | | CCSD/hAWCV{T,Q}Z half-CP | | |
|---|---|---|---|---|---|---|---|---|---|
| | PNO-LMP2 | PNO-LCCSD | (T) | PNO-LCCSD(T) | CCSD-MP2 | CCSD(T)-MP2 | MP2 | CCSD | CCSD-MP2 |
| | | | | $TeO_3 \cdots TeHF$ | | | | | |
| $0.9r_e$ | 2.551 | 1.774 | 0.022 | 1.796 | -0.777 | -0.756 | 2.662 | 1.721 | -0.941 |
| $0.95r_e$ | 0.664 | 0.426 | 0.029 | 0.455 | -0.238 | -0.209 | 0.750 | 0.392 | -0.357 |
| $1.0r_e$ | -0.464 | -0.422 | 0.035 | -0.387 | 0.042 | 0.077 | -0.394 | -0.444 | -0.050 |
| $1.05r_e$ | -1.026 | -0.897 | 0.020 | -0.876 | 0.130 | 0.150 | -0.967 | -0.912 | 0.055 |
| $1.10r_e$ | -1.197 | -1.092 | -0.023 | -1.115 | 0.105 | 0.082 | -1.147 | -1.102 | 0.045 |
| $1.25r_e$ | -0.774 | -0.759 | -0.091 | -0.851 | 0.015 | -0.076 | -0.745 | -0.776 | -0.031 |
| $1.50r_e$ | -0.219 | -0.176 | -0.024 | -0.200 | 0.043 | 0.019 | -0.214 | -0.200 | 0.014 |
| $2.0r_e$ | -0.059 | -0.017 | -0.004 | -0.021 | 0.042 | 0.038 | -0.058 | -0.036 | 0.021 |
| | | | | $SeO_3 \cdots SeHF$ | | | | | |
| $0.9r_e$ | 0.012 | -0.142 | 0.010 | -0.132 | -0.154 | -0.144 | -0.006 | -0.178 | -0.173 |
| $0.95r_e$ | -0.385 | -0.460 | -0.008 | -0.468 | -0.075 | -0.083 | -0.403 | -0.488 | -0.085 |
| $1.0r_e$ | -0.537 | -0.588 | -0.027 | -0.615 | -0.051 | -0.078 | -0.551 | -0.607 | -0.056 |
| $1.05r_e$ | -0.554 | -0.594 | -0.042 | -0.636 | -0.041 | -0.082 | -0.564 | -0.611 | -0.048 |
| $1.10r_e$ | -0.507 | -0.527 | -0.046 | -0.573 | -0.020 | -0.066 | -0.513 | -0.555 | -0.042 |
| $1.25r_e$ | -0.298 | -0.292 | -0.034 | -0.326 | 0.006 | -0.028 | -0.300 | -0.320 | -0.020 |
| $1.50r_e$ | -0.114 | -0.098 | -0.010 | -0.108 | 0.016 | 0.006 | -0.116 | -0.116 | 0.000 |
| $2.0r_e$ | -0.028 | -0.013 | -0.004 | -0.016 | 0.015 | 0.011 | -0.026 | -0.022 | 0.005 |

| | PNO-LCCSD(T)/(tight, haWCV{Q,5}Z) half-CP | | | | | Canonical CCSD/haW{T,Q}Z half-CP | | | |
|---|---|---|---|---|---|---|---|---|---|
| | RMSD | | MAD | | | RMSD | | MAD | |
| | Te | Se | Te | Se | | Te | Se | Te | Se |
| PNO-LMP2 | 1.135 | 0.371 | 0.869 | 0.304 | DF-MP2 | 1.154 | 0.379 | 0.867 | 0.310 |
| PNO-LCCSD | 0.874 | 0.404 | 0.695 | 0.339 | DF-CCSD | 0.866 | 0.424 | 0.698 | 0.362 |
| (T) | 0.039 | 0.027 | 0.031 | 0.022 | | | | | |
| PNO-LCCSD(T) | 0.891 | 0.428 | 0.713 | 0.359 | | | | | |
| CCSD-MP2 | 0.295 | 0.066 | 0.174 | 0.047 | CCSD-MP2 | 0.358 | 0.075 | 0.189 | 0.054 |
| CCSD(T)-MP2 | 0.287 | 0.076 | 0.176 | 0.062 | | | | | |

Furthermore, as discussed earlier in this section, $(n-1)$d effects are repulsive for both equilibrium and stretched geometries, which is somewhat surprising but could be attributed to the systems explored in this study. As expected, these effects gradually diminish for stretched geometries and effectively reduce to very small values at $2.0r_e$. However, for compressed geometries, $(n-1)$d correlation effects are attractive for Te complexes.

Conducting a similar examination on the $(n-1)$sp subvalence correlation reveals that in the case of Te-complexes, the $(n-1)$sp component of (T) is significantly larger than what was observed for $(n-1)$d (0.109 versus 0.040 kcal mol$^{-1}$), and consistently exhibits a repulsive nature. Therefore, although the $(n-1)$sp subvalence component appears to be an order of magnitude smaller than the $(n-1)$d contribution in the total PNO-LCCSD(T) interaction energies, the PNO-LMP2 and (T) corrections operate in opposite directions, resulting in error cancellation at the PNO-LCCSD(T) level. Thus, it is imperative to incorporate $(n-1)$sp subvalence correction as well, particularly when our calculations are confined to the MP2 level.

### 3.3 Spin–orbit coupling

The above observations concerning the importance of subvalence correlation might lead the reader to wonder whether, for such heavy elements as Te, second-order spin–orbit coupling (SOC2) might not be of comparable importance. (As all species considered here are closed-shell singlets, there is no first-order spin–orbit contribution.) We computed SOC2 for each of the 24 chalcogen-bonded dimers using STEOM-CCSD in ORCA, with the aug-cc-pVTZ-DK basis set.

Fig. 1 illustrates the SOC2-induced stabilization of the ground state in relation to the binding energies of all 24 systems at the STEOM-CCSD/aug-cc-pVTZ-DK level. The binding energies in this scenario are calculated as the negative of the difference between the energy of the complex and the energy when the monomers are at a distance where SOC2 effects are negligible (at a sufficiently long distance SOC2 effects become negligible). Additionally, the plot depicts the magnitude of inner-shell subvalence $(n-1)$d and $(n-1)$sp correlation contributions, as discussed in the preceding section, at the PNO-LCCSD(T)(tight, haWCV{Q,5}Z) half-CP level. Here, again, a negative value indicates destabilization effects. For TeO$_3$···TeHF complexes, at equilibrium separation, both SOC2 and the inner-shell subvalence $(n-1)$d contribute to the destabilization of the dimer, with SOC2 and $(n-1)$d having the same magnitude of around $-0.4$ kcal mol$^{-1}$.

Upon compression of the dimer, SOC2 continues to contribute approximately $-0.5$ kcal mol$^{-1}$, but the inner-shell subvalence $(n-1)$d begins to stabilize the TeO$_3$···TeHF dimer, resulting in opposing corrective effects. Conversely, upon stretching the TeO$_3$···TeHF, SOC2 remains constant up to a distance of $1.10r_e$, but the subvalence $(n-1)$d destabilizes by 1.119 kcal mol$^{-1}$. As the Te–Te distance is further extended, both effects diminish. For instance, at a distance of $1.5r_e$, the SOC2 effects become negligible, although the contribution of $(n-1)$d remains slightly significant at $-0.2$ kcal mol$^{-1}$.

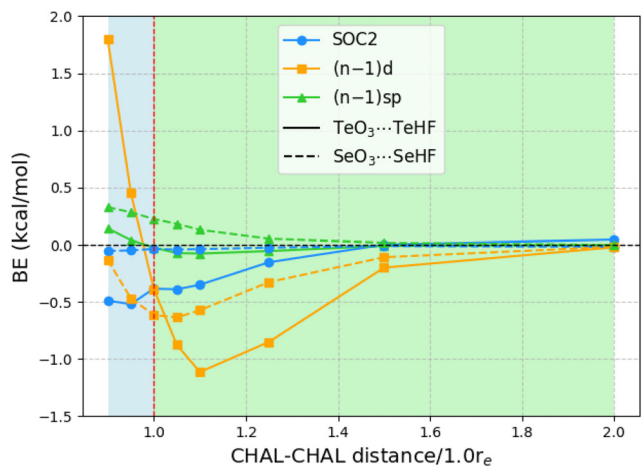

Fig. 1 Impact of second-order spin–orbit coupling (SOC2) on the binding energies of 24 systems calculated at the STEOM-CCSD/aug-cc-pVTZ-DK Level, accompanied by subvalence $(n-1)$d and $(n-1)$sp correlation contributions determined at the PNO-LCCSD(T)(tight, hAWCV{Q,5}Z) half-CP Level. Note that the binding energy (BE) is expressed as the negative of the difference between the complex and monomer energies. For SOC2, this is the negative of the difference between the energy of the complex and the energy when the monomers are at a distance where SOC2 effects are negligible. Thus, positive BE values indicate dimer stabilization.

For the Se complexes, SOC2 effects are much less significant, consistent with the apparent $\propto Z^4$ proportionality of SOC2. (By way of illustration: already in 1998, Runeberg and Pyykkö found that the SOC2 stabilizations of Xe$_2$ and Rn$_2$ are 0.7 and 4.5 meV, respectively—a ratio of $6.4 \approx (86/54)^4$. Similarly, Feller et al. found SOC2 stabilizations of 0.4 and 2.0 kcal mol$^{-1}$, respectively, for Br$_2$ and I$_2$ diatomics, consistent with $(53/35)^4 \approx 5.3$.) For example, at equilibrium distances, SOC2 destabilizes the dimer by only 0.035 kcal mol$^{-1}$, whereas incorporating $(n-1)$d subvalence electrons has an effect of 0.615 kcal mol$^{-1}$ towards destabilization. As we extend the Se–Se distances, both SOC2 and subvalence $(n-1)$d contributions converge to zero. The effect of subvalence $(n-1)$sp remains negligible throughout.

In order to rule out the possibility that the SOC2 we observed above is not a cumulative effect from the two chalcogens in ChO$_3$···ChHF, as one reviewer suggested, we also explored a simpler NH$_3$···ChHF system. The computed values are reported in ESI.† For NH$_3$···TeHF, the SOC2 value is $-0.754$ kcal mol$^{-1}$. Upon compression, the SOC2 continues to destabilize; for instance, the value is $-1.045$ kcal mol$^{-1}$ for $0.90r_e$. When stretching NH$_3$···TeHF, the value remains nearly the same, and as expected, SOC2 diminishes to $-0.011$ kcal mol$^{-1}$ for $2.0r_e$.

For NH$_3$···SeHF, the SOC2 effects are an order of magnitude smaller at $-0.085$ kcal mol$^{-1}$. This value remains nearly the same when compressing or stretching the dimer; for instance, it is $-0.141$ kcal mol$^{-1}$ for $0.90r_e$ and $-0.068$ kcal mol$^{-1}$ for $1.10r_e$.

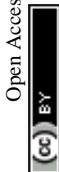

Table 3  Indices delineating noncovalent interaction types and their progression along the dissociation curve for all 24 chalcogen-bonded complexes investigated in this study. Computed interaction energies (kcal mol$^{-1}$) at DF-CCSD/hAWCV{T,Q} and PNO-LCCSD(T)(tight, haWCV{Q,5}Z) half-CP levels are included for clarity

|  | NDF2 | CSPI | DEBC | Hobza ratio | %HF | (A) BE (kcal mol$^{-1}$) | (B) BE (kcal mol$^{-1}$) |
|---|---|---|---|---|---|---|---|
| **TeO$_3$···TeHF** | | | | | | | |
| 0.90$r_e$ | −402.048 | −88.398 | 1.000 | 0.38 | 100.51 | 23.221 | 21.145 |
| 0.95$r_e$ | 34.805 | 8.424 | 0.993 | 0.39 | 96.55 | 30.523 | 29.366 |
| 1.0$r_e$ | 11.801 | 3.146 | 0.953 | 0.40 | 92.59 | 31.278 | 30.936 |
| 1.05$r_e$ | 6.238 | 1.837 | 0.878 | 0.42 | 88.74 | 28.525 | 28.784 |
| 1.10$r_e$ | 4.151 | 1.355 | 0.805 | 0.44 | 85.41 | 24.285 | 24.848 |
| 1.25$r_e$ | 2.581 | 1.098 | 0.739 | 0.51 | 80.05 | 12.701 | 12.955 |
| 1.50$r_e$ | 2.210 | 1.037 | 0.720 | 0.64 | 80.86 | 5.137 | 5.279 |
| 2.0$r_e$ | 4.442 | 1.910 | 0.886 | 0.57 | 93.16 | 1.282 | 1.325 |
| **SeO$_3$···SeHF** | | | | | | | |
| 0.90$r_e$ | 4.308 | 1.095 | 0.738 | 0.27 | 21.51 | 8.844 | 9.513 |
| 0.95$r_e$ | 3.323 | 0.946 | 0.687 | 0.30 | 50.90 | 12.075 | 13.018 |
| 1.0$r_e$ | 2.617 | 0.833 | 0.640 | 0.33 | 57.72 | 12.618 | 13.634 |
| 1.05$r_e$ | 2.132 | 0.754 | 0.602 | 0.37 | 60.15 | 11.887 | 12.846 |
| 1.10$r_e$ | 1.798 | 0.698 | 0.572 | 0.40 | 61.32 | 10.670 | 11.356 |
| 1.25$r_e$ | 1.263 | 0.588 | 0.507 | 0.54 | 63.90 | 7.033 | 7.492 |
| 1.50$r_e$ | 0.993 | 0.476 | 0.430 | 0.74 | 69.85 | 3.321 | 3.577 |
| 2.0$r_e$ | 1.026 | 0.462 | 0.420 | 0.72 | 78.85 | 0.727 | 0.776 |
| **SO$_3$···SHF** | | | | | | | |
| 0.90$r_e$ | 1.647 | 0.437 | 0.401 | 0.29 | −5.56 | 7.897 | 9.328 |
| 0.95$r_e$ | 1.267 | 0.392 | 0.365 | 0.31 | 20.56 | 9.431 | 10.756 |
| 1.0$r_e$ | 0.988 | 0.356 | 0.335 | 0.34 | 31.95 | 9.762 | 10.965 |
| 1.05$r_e$ | 0.791 | 0.329 | 0.312 | 0.37 | 38.80 | 9.456 | 10.528 |
| 1.10$r_e$ | 0.654 | 0.309 | 0.295 | 0.41 | 43.79 | 8.821 | 9.745 |
| 1.25$r_e$ | 0.449 | 0.271 | 0.261 | 0.52 | 54.30 | 6.376 | 7.004 |
| 1.50$r_e$ | 0.401 | 0.254 | 0.247 | 0.66 | 65.34 | 3.149 | 3.349 |
| 2.0$r_e$ | 0.516 | 0.320 | 0.305 | 0.60 | 75.98 | 0.669 | 0.720 |

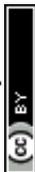

Thus, it's clear that the observed SOC2 effects are not simply due to a cumulative effect from the two chalcogens.

SOC2 values computed using the SOC-TD-DFT code in ORCA at the all-electron CAM-B3LYP/aug-cc-pVQZ-DK level are reported in the ESI.†

### 3.4 Nature of non-covalent chalcogen interactions

Finally, we investigate the nature of non-covalent interactions present in all 24 systems considered in this manuscript. Symmetry-Adapted Perturbation Theory (SAPT, see ref. 87 for a review) provides a systematic framework for breaking down the total interaction energy into distinct components, thus clarifying the nature of intermolecular forces. Following the notation of ref. 69 and 88 at the two least expensive levels of SAPT, namely SAPT0 and SAPT2, the interaction energy can be partitioned as follows:

$$IE_{SAPT0} = E_{elst}^{(10)} + E_{exch}^{(10)} + E_{ind}^{(20)} + E_{exch-ind}^{(20)} + E_{disp}^{(20)} + E_{exch-disp}^{(20)} \quad (1)$$

$$IE_{SAPT2} = IE_{SAPT0} + E_{elst}^{(12)} + E_{exch-elst}^{(12)} + E_{ind}^{(22)} + E_{exch-ind}^{(22)} \quad (2)$$

where the terms colored in blue denote attractive forces, those in red represent repulsive forces, while terms in black may exhibit either attractive or repulsive behavior. The two superscripts indicate the order of intermolecular and intramolecular perturbation theory, respectively, while the subscripts "exch", "elst", and "ind", and correspond to exchange repulsion, electrostatic interaction, and induction, respectively. Our SAPT findings, calculated using the def2-QZVPP basis set, are documented in the ESI.†

Before delving into our results, let us consider various indices that have been developed to identify the nature of non-covalent interactions. The Hobza dispersion/electrostatic ratio[98] uses the information from the above equations: $D/E \geq 1.7$ is deemed dispersion-dominant, $D/E \leq 0.59$ (i.e., 1/1.7) electrostatic-dominant, and the range in between 'mixed-influence'.

One of us has proposed alternative indices (see ref. 69) that only entail MP2 calculations and do not necessitate SAPT computations. The first of these approaches is the correlation spin polarization index (CSPI), which serves as an indicator of the nature of non-covalent interactions.

$$CSPI = \frac{IE_{ss}^{(2)} - IE_{ab}^{(2)}}{IE_{ss}^{(2)} + IE_{ab}^{(2)}} \quad (3)$$

In systems where the interaction energy is primarily governed by dispersion forces, CSPI tends to approach zero. However, in systems where non-dispersion factors contribute to the correlation portion of the interaction energy, CSPI will deviate significantly from zero.

Nevertheless, in the case of nearly dissociated dimers, the absolute values of IE$_{aa}$ and IE$_{ab}$ become so small that the CSPI

may change sign. To address this issue, the authors have also introduced the DEBC index (dispersion–electrostatic balance in correlation).

$$\text{DEBC} = \sqrt{\frac{\text{CSPI}^2}{1 + \text{CSPI}^2}} \qquad (4)$$

DEBC spans a scale from 0, indicating a system dominated purely by dispersive forces, to 1, indicating a system governed solely by non-dispersive interactions. This should be considered in tandem with %HF (percentage of Hartree–Fock in the interaction energy), which is defined as follows:

$$\%\text{HF} = \frac{(100\%)\text{IE}_{\text{SCF}}}{\text{IE}_{\text{SCF}} + \text{IE}_{\text{aa}}^{(2)} + \text{IE}_{\text{bb}}^{(2)}} \qquad (5)$$

The %HF value tends towards 100% in systems where binding primarily results from pure electrostatic effects such as dipole–dipole interactions.

Returning now to the 24 chalcogen bonded systems investigated in this paper, Table 3 displays the NDF2, CSPI, DEBC, Hobza ratio (dispersion/electrostatic), and %HF for all 24 chalcogen bonded dimers. Additionally, interaction energies computed using PNO-LCCSD(T)/(tight, haWCV{Q,5}Z) half-CP and DF-CCSD/hAWCV{T,Q}Z half-CP are included in the last two columns.

At equilibrium distances, the Hobza ratios for the $\text{TeO}_3\cdots\text{TeHF}$, $\text{SeO}_3\cdots\text{SeHF}$, and $\text{SO}_3\cdots\text{SHF}$ complexes are 0.40, 0.33, and 0.34, respectively. These values suggest that the interactions lean towards the electrostatic-dominated end of the spectrum, albeit to a lesser extent than a purely hydrogen-bonded complex like the acetic acid dimer, for which the value is around 0.5, as reported in Table 16 of the ref. 69. %HF also indicates that Te complexes are dominated by electrostatic effects, followed by Se and S, a trend consistent with CSPI and DEBC analyses.

Moreover, in all three complexes, even the stretched geometries (e.g.; $2.0r_e$) persist within the electrostatic realm (more towards the 'mixed-influence' region) as indicated by the Hobza ratios of 0.57, 0.72, 0.60, respectively, for the Te, Se and S complexes. The respective %HF values are 93%, 79%, and 76% respectively.

Therefore, although definitive conclusions regarding the nature of NCIs investigated in this study are challenging to draw, it is evident that dispersion effects are minimal across all 24 complexes examined.

## 4 Conclusions

Our computational investigation into chalcogen bonding interactions has led us to conclude the following:

1. The inclusion of inner-shell $(n - 1)$d subvalence correlation destabilizes chalcogen dimers by nontrivial amounts—a trend opposite to that observed for halogen bonding interactions in ref. 37. (At highly compressed geometries, however, subvalence $(n - 1)$d correlation begins to stabilize the complex.) Among the various interaction energies components computed at the PNO-LCCSD(T) or DF-CCSD levels, the PNO-LMP2 or DF-MP2 component of the $(n - 1)$d correlation constitutes the lion's share, which could prove useful in future investigations. Keeping the $(n - 1)$d correlation treatment down to the MP2 level affords significant savings in CPU time and memory/storage overhead.

2. The effect of the lower $(n - 1)$sp subvalence correlation is notably less pronounced. The PNO-LMP2 and (T) components exert opposing influences, leading to error cancellation at the PNO-LCCSD(T) level. Therefore, if one is limited to MP2 level calculations, it is essential to also include $(n - 1)$sp subvalence correlation.

3. The SOC2 effects appear to be less significant than the $(n - 1)$d correlation; however, they remain non-trivial, particularly for Te complexes. For the Se complexes, SOC2 effects are much less significant, consistent with the approximate $\propto Z^4$ proportionality of SOC2. SOC2 stabilizes the monomer more than the dimer (as the dimer orbitals are separated further and hence the denominator in the interaction is reduced), and as a result destabilizes the dimers. At equilibrium and stretched geometries, SOC2 and $(n - 1)$d destabilize the complex in tandem; at compressed geometries, however, they work in opposite directions as $(n - 1)$d becomes stabilizing there.

4. We observe that all three complexes, $\text{TeO}_3\cdots\text{TeHF}$, $\text{SeO}_3\cdots\text{SeHF}$, and $\text{SO}_3\cdots\text{SHF}$, tend towards the electrostatic-dominated end of the spectrum at shorter, but enter a mixed-influence regime at longer, intermonomer separations.

5. Employing complete basis set (CBS) extrapolation along with CP correction appears to be the most effective strategy. Interestingly, {T,Q}Z extrapolation outperforms the 5Z basis set, highlighting its potential superiority in these analyses. Additionally, we observe a slower-than-usual convergence of basis sets for the SCF component.

## Data availability

All data supporting the findings of this study are included within the article and its ESI.†

## Conflicts of interest

The authors declare that there is no conflict of interest.

## Acknowledgements

Work on this paper was supported by the Israel Science Foundation (grant 1969/20) and by a research grant from the Artificial Intelligence and Smart Materials Research Fund (in memory of Dr Uriel Arnon), Israel. Nisha Mehta would like to acknowledge the Feinberg Graduate School for a Sir Charles Clore Postdoctoral Fellowship as well as Dean of the Faculty and Weizmann Postdoctoral Excellence fellowships. Nisha Mehta acknowledges the School of Chemistry, The University of Melbourne, for the Visiting Academic appointment. Nisha Mehta would like to extend her heartfelt thanks to her friend,

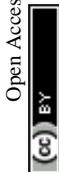

Dr Marcos Casanova-Páez, for introducing her to STEOM-CCSD in ORCA.

# Notes and references

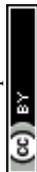